# Quantum-Resilient Banking Transaction Security using DW-HKEM: A Hybrid RSA/ML-KEM Cryptographic Gateway with Real-Time Monitoring

**Siddhaarth S Prabhu**[1] **Aswani Kumar Cherukuri** [1,2]

[1]School of Computer Science Engineering and Information Systems

[1,2]Centre For Artificial Intelligence Research (C-FAIR)

Vellore Institute of Technology, Vellore 632014, India

[1]sidhu1401.sp@gmail.com, [2]cherukuri@acm.org

**ABSTRACT**

The increasing number of online banking and financial services on public-facing networks has caused security of cryptographic protocols to become a serious, systemic problem. RSA-2048 is the key length that is used in the majority of transactions that travel over the Internet every day, including fund transfers, authentication exchanges and key negotiations for new sessions. A very insidious threat vector, the Harvest Now, Decrypt Later (HNDL) paradigm, is already in existence with the adversarial nation-state actors and well-funded threat actors who spy on and archive encrypted banking traffic today with the clear goal of decrypting it later in the day when cryptographically relevant quantum computers (CRQCs) arrive. The exposure window is not a time in the future – it is the NOW moment! Quantum computation is a certain threat to the current public-key cryptographic infrastructure, and one that is moving ever closer. At the same time, an unheralded weakness still exists in the deployed financial systems: operational obscurity of cryptographic operations (OcOps), in which the encryption and decryption processes continue as a common black-box operations that lack real-time observability, and that prevents the security architect from identifying algorithm downgrade attacks, resource exhaustion, and compromised cryptographic libraries. In this paper, the authors take up both of these challenges, designing, implementing and testing an integrated system based on a Dual-Wrap Hybrid Key Encapsulation Mechanism (DW-HKEM) and a live cryptographic analytics dashboard. Both the RSA-256 and ML-KEM-768 are hosted within the DW-HKEM framework and an AES-256 session key is wrapped in each of them, with a composite key created using a Key Derivation Function based on SHA-256, resulting in combinatorial security against classical and quantum adversaries and backwards compatibility with existing banking systems. The CryptoX real-time dashboard offers time tracking for each operation, session-level analytics and visualization of performance trends for immediate use in enterprise banking security operations. The results from empirical benchmarking across 100 independent iterations show a total overhead of about 1.58ms, which is well within the enterprise deployment threshold, proving the architecture viable as a practical, scalable reference for the financial-sector's cryptographic infrastructure and applications to move to a post-quantum era. We also provide complete source code of the work carried out.

Keywords: Post-Quantum Cryptography, Hybrid Encryption, ML-KEM-768, RSA-2048, AES-256, Harvest Now Decrypt Later, Banking Transaction Security, Key Encapsulation, Cryptographic Analytics, HNDL, Financial Security, FIPS 203.

## I. INTRODUCTION

The bearer of the present-day digital infrastructure is cryptography. The mathematical computational infeasibility of problems that are impossible to solve on classical computers is the basis for all secure channels, authenticated transactions and digitally signed documents. The most widely used asymmetric cryptosystem, RSA-2048, is based on the integer factorization problem: When n is a product of two large primes p and q, it is known that p and q can be found by an exponential amount of classical (non-quantum) computation. This mathematical hardness is the foundation of TLS, SSH, PKI, S/MIME and a wide range of enterprise security protocols that have been in use for nearly 50 years and are used worldwide to establish digital trust. This reliance is more pronounced in the financial services industry. RSA is used in every online banking site, real-time payment gateways, interbank settlement networks and mobile transaction application for secure session channel between the customer and financial servers. The session negotiation (including symmetric key establishment) is protected by protocol-level RSA encryption when consumers initiate a session with a public network, such as for a wire transfer or a credit card authorization or an account authentication. The reliance is not just a technical quirk, but is deeply rooted in the regulatory compliance frameworks, hardware security modules and the certificate chains in which financial services institutions rely as they run the cryptographic operations, they rely on worldwide.

In a quantifiable and nearby way, however, this architecture is quantum-vulnerable. It was shown by Peter Shor in 1994 that a sufficiently powerful quantum computer can factor numbers of size N in polynomial time [15]. There are signs that practical implementation of cryptographically-relevant quantum computers is near, although they have not yet left the "noisy intermediate scale quantum" (NISQ) hardware development era, characterized by the fact that qubit scaling and error correction techniques developed by IBM, Google and national research labs are steadily progressing. For financial institutions, the window of opportunity for doing the migration in a cryptographic manner is closing.

The Harvest Now, Decrypt Later (HNDL) attack paradigm is a big risk factor in this case. The ability to intercept and archive encrypted network traffic has been scaled up and is now being done by nation-state adversaries and high-end threat actors. Banking transaction packets, such as session handshakes, credentials exchanged in the encrypted form, and payment authorizations are vulnerable to passive interception and can be accomplished without disrupting the connection. The attacker doesn't have to crack the traffic today, all they have to do is record it. After the creation of a CRQC, the whole financial traffic that is collected and encrypted by the RSA can be rendered vulnerable retroactively [3][5]. The implication here is that the banking-omniscience exposure window isn't in the future, sensitive financial data sent now via RSA is being harvested for future decryption.

Another vulnerability, also critical but not as obvious, is the way cryptographic systems are used when deployed in financial institutions. In almost all production setting, the encryption and decryption processes are black-box ones. Security operations teams are not able to observe in real time the number of cryptographic operations executed in a specific time frame, whether they are becoming longer than their baseline (which could be a sign of algorithm downgrade attack or of library compromise) or the operational cost of switching to a new family of algorithms in terms of concrete, measurable performance. This lack of cryptographic observability leaves a room for the security incident response to be compromised, and for the rational planning of post quantum migration time tables to be compromised.

The National Institute of Standards and Technology (NIST) has finished a long-term Post-Quantum Cryptography (PQC) Standardization Project, and issued its initial post-quantum standards in 2024 to overcome the quantum standardization hurdle. The major standard, FIPS 203, establishes the Module-Lattice-Based Key Encapsulation Mechanism (ML-KEM) based on the classical and quantum intractability of a structured lattice problem known as Module Learning With Errors (MLWE) [1][8] which is conjectured hard for both classical and quantum attackers. The need to move to ML-KEM or a compatible hybrid construction in the financial infrastructure is from an academic point of view, but now a regulatory and operational requirement.

Both of these challenges are tackled in this paper by a combination of architecture. Encapsulated under RSA-2048 and ML-KEM-768, the Dual-Wrap Hybrid Key Encapsulation Mechanism (DW-HKEM) combines these two mechanisms into a single protection mechanism, which is backward compatible with the existing banking system relying on RSA, with an added layer of protection based on a lattice function is considered quantum safe. The CryptoX real-time dashboard delivers real-time cryptographic performance observability that can be directly applied to the financial sector security operations. These systems are the building blocks of a deployable Reference Architecture for enterprise banking to migrate to Post Quantum.

### *A. Real-Time Banking Threat Scenario*

Now imagine the following threat scenario—made up of known attack models found in the cybersecurity literature and intelligence community reports—that is grounded in the reality of operations.

The retail banking customer (hereafter referred to as the Transaction Principal (TP)) logs in to the bank's Internet website in a regular browser session on a public Internet connection to transfer funds online via the web portal for purchase of INR 5,00,000 worth of goods. The browser initiates TLS 1.2 with the application server of the bank using a key exchange over RSA-2048, to negotiate a symmetric session key. The result is a symmetric key which is then used to encrypt the transaction payload (account numbers, amounts, authentication tokens) before it is sent. Both the TP and bank are secure with the transaction – the channel is encrypted, the certificate is valid, there is no connection anomaly. The whole session will be concluded normally and the transaction will be posted.

At the same time, however, a network adversary (which is sitting in the middle of the Internet, at an Internet exchange point, on a compromised BGP router or at a rogue access point in the TP's network path) is able to take a copy of the entire encrypted TLS session packets. The adversary does not try and decrypt the message, the protection of the RSA-2048 is not feasible with classical means. Instead, the encrypted session dump – the negotiation of the session key by RSA (encrypted), the encrypted payload of the transaction, and all the metadata – is saved inside an encrypted archive data base, which is held by the adversary's infrastructure. The capture operation requires less than 50 mS of network time, and does not produce any anomaly of any magnitude in the bank's security monitoring. The bank's security operations team doesn't even realize that it has been exfiltrated.

This is the attack model used in this HNDL. The cost of the opponent's investment is low today: passive interception infrastructure, storage of a large amount of data, and waiting. When the CRQC is issued, the session key (along with other information like account numbers and balances) is stored in a polynomial-time factorable RSA-encrypted state, and can be retrieved by Shor's algorithm and decrypted to provide the original information (account numbers, balances, authentication material). If the financial data still has value at that time – whether as an account credential, historical transaction intelligence or for regulatory purposes, the adversary has made his/her investment pay off. There isn't anything that the TP and the bank can do: the data was collected from public infrastructure, and at the time it was collected there was no obligation to notify of the breach.

In this situation there are two serious shortcomings of the traditional banking cryptographic architecture. First, the security guarantee of the RSA-2048 is difficult to achieve in terms of time, as it's dependent on the advances of quantum hardware — a time that is unknown, and not under the financial institution's control. Secondly, any traffic monitoring system that would be able to correlate traffic patterns, identify abnormal captures of sessions and flag cryptographic performance deviations is missing, and therefore the institution is unable to assess its HNDL exposure. It has no idea how much of its transaction traffic it is harvesting, what customers are being affected and if its cryptographic library is operating as expected.

In this paper, it is proposed a DW-HKEM system to directly tackle both failure modes. Encapsulate each session key under both RSA-2048 and ML-KEM-768: If an HNDL adversary captures some transaction session key and then decrypts it by Shor's algorithm on a future CRQC, it gets only one half of the KDF input (namely S_RSA). The composite session key K_Session = SHA-256(S_RSA || S_KEM) is as long as ML-KEM-768 layer is computationally secure as long as the MLWE hardness assumption stands. Today's adversary, who is able to harvest the DW-HKEM-protected banking traffic, is only able to get one bit of the two-component KDF input, which is not enough to retrieve the session key or the transaction payload. Conventional black-box banking systems don't have this capability for security operations to identify performance deviations, like unexpected execution time increases, which can signal a compromise of library or downgrade attack, something the CryptoX monitoring dashboard offers.

## II. BACKGROUND

### *A. RSA-2048 and Its Limitations*

One of the most widely-used public-key cryptosystems used in TLS and PKI today, and for secure banking transactions, is RSA, which was introduced by Rivest, Shamir and Adleman in 1978 [12]. Although it has been widely used, Shor showed that in the presence of powerful quantum computers, it was possible to factor large integers in polynomial time, posing a threat to the security of RSA in the long term [15]. To overcome this problem, this work combines both RSA-2048 and ML-KEM-768 into a hybrid DW-HKEM framework.. For financial-grade TLS deployments around the world, the de facto standard for establishing a TLS session key is RSA-2048. One of its main weaknesses for applications in banking is the computational latency in generating RSA-2048 keys, which, in empirical testing, took about 103.96ms to generate; this is a worthwhile time investment if the key is going to be rotated frequently or new keys generated on the fly. These two are poor bases on which to build a long-term bank cryptographic infrastructure.

### *B. Post-Quantum Cryptography: ML-KEM-768*

In 2024, NIST finalized the ML-KEM as a quantum resistant key encapsulation mechanism in FIPS 203.In 2024, NIST approved ML-KEM as a quantum resistant key encapsulation mechanism in FIPS 203. Avanzi et al. explained how the design of the CRYSTALS-Kyber works and showed its applicability for post-quantum key exchange [8]. These works are, however, mostly aimed at algorithm design and standardization, and not integration with existing banking systems. Hence in this work, the authors used the ML-KEM-768 in addition to RSA-2048 in a hybrid banking security architecture.The main compromise is between the size of the key and the size of the ciphertext: ML-KEM-768 requires 1,184 bytes for public key and 1,088 bytes for ciphertext while RSA-2048 uses 256 bytes for both. The Open Quantum Safe (OQS) project is currently developing a high-performance C implementation of liboqs, which is based on the Number Theoretic Transform (NTT) for efficient polynomial arithmetic [2].

### *C. AES-256 and Symmetric Encryption*

Advanced Encryption Standard with 256-bit keys (AES-256) is a symmetric block cipher standardized by the NIST (FIPS 197 [14]). It has been designed to work on 128-bit blocks with 256 bits of classical security. Theoretically quantum computers can search the key space of AES-256 in $O(2^{128})$ operations, which is still impractical, so this is quantum-resilient in practice. The AES-256-CBC (Cipher Block Chaining) is used when encrypting messages, and a random 128-bit Initialization Vector (IV) is added to the beginning of the ciphertext in the DW-HKEM construction. AES-256-GCM is used for the hybrid encryption layer in CryptoX web system, ensuring both confidentiality and authentication. The rationale and construction of hybrid cryptography are explained. The concept and design of hybrid cryptography is discussed.

### *D. Hybrid Cryptography: Rationale and Construction*

The hybrid model, which combines classical and post-quantum algorithms, represents the best approach to the transition between the classical and quantum cryptographic periods and is especially suitable for the implementation in financial services. The rationale involves 3 things. First, this is because ML-KEM has not been cryptanalysed in the same depth or with the same degree of scrutiny as has RSA ever been. A hybrid construction also means that if an MLWE is broken in an unlikely way, it will not break all financial sessions. Secondly, due to the fact that RSA is tightly coupled with legacy banking hardware (HSMs, X.509 PKI chains), wholesale protocol replacement is simply not doable without the use of a hybrid transitional architecture. Third, the additional computational cost of hybrid operation (~1.58ms per (full) handshake) is unnoticeable for enterprise deployments, TLS 1.3 handshake budgets [7].

Table 1 shows the comparison between RSA-2048, ML-KEM-768, and proposed DW-HKEM in terms of Latency, Key Size and Quantum Resistant.

***Table I: Comparative Analysis — RSA-2048 vs. ML-KEM-768 vs. DW-HKEM Hybrid***

| Parameter | RSA-2048 | ML-KEM-768 | DW-HKEM (Hybrid) |
|---|---|---|---|
| Security Basis | Integer Factorization | MLWE (Lattice) | Both (Combinatorial) |
| Quantum Resistance | Broken by Shor's Alg. | Conjectured Secure | Secure (MLWE Layer) |
| Key Gen Latency | 103.96 ms | 0.63 ms | ~104.58 ms |
| Encapsulation Latency | 0.38 ms | 0.08 ms | 0.46 ms |
| Decapsulation Latency | 8.20 ms | 0.10 ms | 8.30 ms |
| Public Key Size | 256 B | 1,184 B | 1,440 B |

| Parameter | RSA-2048 | ML-KEM-768 | DW-HKEM (Hybrid) |
|---|---|---|---|
| Ciphertext Size | 256 B | 1,088 B | 1,392 B |
| NIST Status | Legacy (FIPS 186) | FIPS 203 (2024) | FIPS 203 Compliant |

### *E. Real-Time Scenario*

The banking systems of today process thousands of transactions online per second, such as payments and transfers, logging in, and the movement of sensitive data among their customers. The vast number of these communications still rely on conventional encryption of the data being sent using RSA. The issue is an attacker can still steal and record this encrypted banking traffic through the process of packet interception, public Wi-Fi attacks or Man-in-the-Middle (MITM) attacks. Even though it is not possible to decrypt the data yet, the quantum computers of the future might be able to use Shor's algorithm to break the RSA encryption. This creates a very serious 'Harvest Now Decrypt Later' threat as it means that financial data that has been encrypted today may be decrypted in the future. In addition to this, there is no real-time visibility on cryptographic operations in most of the existing banking security solutions.

Typical encryption processes are black-box processes, and organizations are often challenged to observe their performance problems, identify anomalies or assess their migration readiness towards post-quantum cryptography. To resolve these issues, the proposed DW-HKEM framework is based on the hybrid of RSA-2048, ML-KEM-768 and AES-256 session security, which can protect banking transactions against classical and future quantum attacks. It also includes a real-time monitoring dashboard with live analytics on cryptographic operations, allowing organizations to keep a close eye on their operations and ensure security and visibility throughout the secure banking communication process.

## RESEARCH CONTRIBUTION

1) Designed and developed a Dual-Wrap Hybrid Key Encapsulation Mechanism (DW-HKEM) based on RSA-2048 and ML-KEM-768 to enhance security for banking transactions.
2) Improved a hybrid KDF (hybrid RSA and ML-KEM shared secrets) to give 2-layer protection of session-key in a SHA-256.
3) Developed and created a real-time cryptographic monitoring dashboard with ability to monitor encryption, decryption and latency metrics.
4) Carried out 100-iteration benchmark tests to measure computational and communication cost of hybrid post-quantum deployments.
5) Compared the suggested design with the Harvest-Now-Decrypt-Later (HNDL) threat model that applies to the banking sector.

## III. LITERATURE SURVEY

***Table II: Literature Survey — Post-Quantum and Hybrid Cryptography Research***

| Ref. | Author(s) | Key Contribution | Limitation |
|---|---|---|---|
| [1] | NIST (2024) | Standardized ML-KEM (FIPS 203) as the primary post-quantum KEM, defining parameter sets 512, 768, | Specification only; no hybrid integration or implementation guidance |

| Ref. | Author(s) | Key Contribution | Limitation |
|---|---|---|---|
| | | and 1024. | provided. |
| [2] | Open Quantum Safe | liboqs: open-source C library implementing all NIST PQC finalists including ML-KEM, with oqs.dll for Windows deployment. | Library-level only; no orchestration, benchmarking, or hybrid key derivation logic. |
| [3] | Bagla et al. (2023) | Evaluated lattice-based cryptography for IoT and cloud security; identified energy and bandwidth as primary deployment obstacles. | No hybrid RSA+PQC implementation; no empirical hybrid overhead measurement. |
| [4] | M. D et al. (2025) | Proposed dual-layer RSA encryption with Merkle Trees for quantum resistance and data integrity. | Uses modified RSA, not NIST-standardized ML-KEM; no empirical benchmarking. |
| [5] | Ambika et al. (2024) | Investigated quantum computing's impact on TLS handshake protocols; showed RSA key exchange is the most vulnerable component. | Focused on theoretical protocol analysis; no post-quantum replacement implemented. |
| [6] | Yang et al. (2022) | Proposed hybrid cryptosystem combining RSA with optical (Fresnel lens) encryption. | Optical layer impractical for software deployment; no PQC primitive integration. |
| [7] | Kumagai et al. (2024) | Examined RSA vulnerabilities through lattice-based cryptanalysis on small exponent keys. | Attack analysis only; no defensive hybrid construction or performance evaluation. |
| [8] | Avanzi et al. (2021) | CRYSTALS-Kyber algorithm specification — the technical basis for ML-KEM. | Specification document; no hybrid integration or Python implementation. |
| [10] | Bernstein & Lange (2017) | Landmark survey of the post-quantum cryptography landscape; classified threats. | Pre-standardization; no concrete implementation benchmarks. |

A literature survey of the recent research works related to post-quantum cryptography and hybrid cryptographic systems is given in Table II. The comparison outlines some of the most significant features and weaknesses of the current methods such as the ML-KEM standardization, hybrid RSA-PQC models, TLS security analysis, and difficulties in implementing post-quantum cryptography.

***Research Gap Analysis***

As shown in Table II, the literature reviewed has been quite impressive with respect to advances in post-quantum cryptography, hybrid cryptographic architectures, and quantum-threat assessment. FIPS 203 was used to standardize ML-KEM by NIST, and Open Quantum Safe project offers practical implementation of post-quantum algorithms [1][2]. The lattice-based cryptography for secure communications, the hybrid cryptographic approach, and the effect of quantum computing on the current security protocols are the subjects of other research interests [3][5][6][8].

The literature surveyed, however, is mostly concerned with the standardization of algorithms, the analysis of protocols or with theoretical security assessment. There has been little work on embedding RSA-2048 and the ML-KEM-768 in a deployable, banking-centric hybrid key encapsulation system. Moreover, existing studies typically do not integrate post-quantum cryptographic security with the ability to observe and monitor real-time operations.

A second gap found in the literature is the absence of an empirical assessment of how a hybrid cryptography solution is deployed in a banking context in a Harvest-Now-Decrypt-Later (HNDL) threat model. Although some papers have been written on threats from quantum computers and migration strategies, few have offered any implementation level benchmarking and operational analytics to aid migration planning and security monitoring.

This study introduces and analyzes a Dual-Wrap Hybrid Key Encapsulation Mechanism (DW-HKEM) combining both RSA-2048 and ML-KEM-768, and implements it using CryptoX, a real-time monitoring platform. The proposed approach is designed to offer quantum-resistant session key security and visibility for financial applications.

## IV. PROPOSED METHODOLOGY

### *A. System Architecture Overview*

The two interoperating subsystems of the integrated system are explicitly designed with the constraint of deployment use for financial sector. The DW-HKEM cryptographic gateway is a modular system in Python language that is designed as a command line application, features the session key encapsulation core, while the CryptoX Flask-based web application offers real-time encryption services and live analytics. Both subsystems use a common set of cryptographic primitives and have excellent modularity: each of the cryptographic elements is implemented in a separate, testable module, and the analytics part of the subsystem is separated from the cryptographic logic to avoid the overhead of monitoring affecting the measurement of underlying performance.

From a security perspective, the DW-HKEM gateway is located at the session initiation layer; that is, the cryptographic handshake that sets up the AES-256 session key that would then be used to protect the transaction payload that would follow. The DW-HKEM gateway will not use TLS-based banking deployments to use only RSA-2048 for this session key encapsulation, but rather will use a dual encapsulation: RSA-2048 OAEP (to ensure backward compatibility with existing bank PKI infrastructure) and ML-KEM-768 (to give quantum resistant key establishment). Even if the RSA component is compromised by a future CRQC the session key K_Session is still secure under the MLWE assumption as the transaction payload is protected by the derived session key K_Session.

The DW-HKEM system is based on a 5-layer service architecture. The first layer (Data Layer) is for the inputting of data and exporting of data to output. Layer 2 (Cryptographic Service Layer) contains four Python modules (rsa_module.py, pqc_module.py, aes_module.py and hybrid_module.py) that contain a cryptographic primitive. The complete lifecycle of cryptographic session operations, such as module invocations and benchmark loops are controlled by Layer 3 (Orchestration Layer, controller.py). Layer 4 has sub-millisecond timing in time. perf_counter() and atomic CSV writes (metrics_module.py). This is the high-level C implementation of the lattice polynomial arithmetic used in the Native Engine Layer (layer 5, oqs.dll).

CryptoX (web system) is client server architecture based. The frontend is a single page HTML/JavaScript application which communicates asynchronously with the Flask backend through the REST/HTTP-JSON. First, the frontend displays two graphs based on Chart.js: a live line plot showing the trends of the execution time (in seconds) for both the RSA and hybrid techniques over operations; a panel showing the total number of operations performed and the average execution time for each operation (Fig. 1). There are four endpoints on the backend: /api/encrypt, /api/decrypt, /api/metrics and /api/reset. The CryptoX dashboard would offer the financial security operations team with real-time visibility to detect performance anomalies, validate migration SLA adherence, and help with cryptographic agility choices.

### *B. Hybrid Encryption Flow for Banking Transactions*

If used in a banking transaction session, the encryption flow would be as follows: When a transaction payload (account identifiers, transaction amounts, and authentication material) is received by the DW-HKEM gateway, the gateway creates a cryptographically random 256-bit AES master key M by calling os.urandom(32) and a random 128-bit Initialization Vector. The RSA public key P with OAEP padding (SHA-256 hash and MGF1 mask generation) is used to encrypt M into a 256-byte RSA ciphertext C_RSA, by applying RSA encryption.RSA encryption is applied using the public key P to encrypt M into C_RSA, an RSA ciphertext of length 256 bytes under OAEP padding (SHA-256 hash, MGF1 mask generation). At the same time, ML-KEM is invoked as a native engine (called oqs.dll), and outputs ML-KEM shared secret S_KEM (256 bits) and ML-KEM ciphertext C_KEM (1,088 bytes). This hybrid session key is then used to create an AES-256-CBC encrypted session payload using K_Session, which is created by the KDF. This means that decrypting a transaction set in the bundle will give an HNDL attacker no benefit as they must be able to break two distinct ciphers: RSA and MLWE; a computational impossibility in any practical timeline.

### *C. Key Derivation Function (KDF) Design*

The KDF is the cryptographic center of the DW-HKEM construction along with the way that HNDL resistance is formally based. Two 256-bit secrets are released after both of the encapsulation operations: S_RSA (AES master key which can be recovered by performing RSA decapsulation) and S_KEM (shared secret of the ML-KEM). These are concatenated and hashed:

$$K_Session = SHA\text{-}256(S_RSA \,||\, S_KEM)$$

SHA-256 is assumed to be a random oracle. A 256-bit K_Session is then computed which is computationally indistinguishable from a uniformly random string, provided that at least one of S_RSA or S_KEM is computationally secure. If the adversary is HNDL that has a CRQC, then against it S_RSA is recovered by Shor's algorithm but not S_KEM (protected by the MLWE assumption). It is therefore computationally secure and the banking transaction payload protected is HNDL-resistant. In the face of a typical opponent who is taking advantage of a potential future vulnerability in the MLWE: S_RSA (protected by the RSA) continues to be secure, acting as the security backstop. It is a security guarantee that is provided in both directions—hence the double wrap construction.

### *D. Real-Time Monitoring and Analytics System*

The CryptoX monitoring system can be used in financial sector security operations directly. For every operation in the system, a log entry is added with the operation type (encrypt/decrypt), execution time (in milliseconds) and Unix timestamp. This log has three different operational capabilities in a deployment scenario for banking:

The execution time trend chart is the first anomaly detection as deviations are easily noticeable, whether due to resource usage, algorithm downgrade attack, or tampered cryptographic library. If, for example, the decapsulation latency for RSA suddenly increases, it may be a problem with the libraries or a slowdown attack on session key recovery.

Second, capacity planning: operation count trends can help security architects decide on important operation rate, scale worker processes and size session key caches based on the peak transaction volume, which is vital for financial institutions that have a defined transaction processing SLA.

Thirdly, migration validation: the concurrent migration timing and the hybrid timing data shown in the Fig. 1 allows banks to validate the hybrid timing in practice with their security teams, before deciding to go all the way in migration to hybrid. Legacy banking cryptographic systems (a class of systems that are “black-box”) are tackled at low engineering cost in order to resolve the observability gap.

### *E. Step-by-Step DW-HKEM Protocol*

The entire DW-HKEM cryptographic protocol is used in a banking transaction session like this:

Step 1 — This operation takes the following times on average: The average time for this operation is: Generation of key pair using lattice arithmetic with oqs.dll from NTT based algorithm (average: 0.63ms). Generates AES master key, master key IV using os.urandom.

Step 2 — AES master key M is then encrypted under RSA public key (RSA-PUB) with OAEP-SHA256, which results in AES master key C_RSA (256 B). Average latency: 0.38 ms.

Step 3 — ML-KEM Encapsulation: kem.encap_secret(pk_pqc) was called via liboqs-python which returns C_KEM (1,088 B) and S_KEM (32 B). Average latency: 0.08 ms.

Step 4 — KDF Derivation: K_Session = SHA-256(S_RSA ‖ S_KEM) done through hashlib.sha256() function. Cryptographically instantaneous (< 0.01 ms).

Step 5 — AES-256-CBC Encryption: Financial information in transit secured by AES-256-CBC encryption of the transaction payload, P, that uses K_Session, the IV and PKCS#7 padding.

Step 6 — Bundle Assembly: Complete ciphertext bundle assembled: C_RSA (256 B) + C_KEM (1,088 B) + IV (16 B) + C_AES. A 2-byte message takes a total of 1,392 B of bundles.

Step 7 — Metrics Logging: Metrics are stored in a file called benchmark.csv, using a strategy of temporary-file-then-rename for Windows and atomically for the rest of the platforms, avoiding PermissionError exceptions for concurrent file access on Windows 11 x86-64.

## V. EXPERIMENTAL RESULTS

### *A. Experimental Environment*

The DW-HKEM benchmarks were run on a x86-64 workstation running Windows 11 (64-bit) and equipped with Python 3.12.2, liboqs v0.10.0. The benchmark program ran 100 random runs of the entire DW-HKEM handshake, using different key pairs each time to avoid key-caching artifacts. The CryptoX dashboard has been tested in real browsers with the Flask development server and Chart.js rendering. All timing was done with the precision of time. perf_counter (), which is sub millisec. All timing is done with time.perf_counter (), which is sub millisec. These conditions are typical of a single instance financial application server and represent the minimum performance achievable in a dedicated-hardware security module, production environment.

### *B. Computational Latency Analysis*

***Table III: Comparative Latency Benchmarking Results (100-Iteration Averages)***

| Algorithm / Phase | Key Gen (ms) | Encapsulation (ms) | Decapsulation (ms) | Total (ms) |
|---|---|---|---|---|
| RSA-2048 (Baseline) | 103.96 | 0.38 | 8.20 | ~112.54 |
| ML-KEM-768 (PQC Only) | 0.63 | 0.08 | 0.10 | ~0.81 |
| DW-HKEM (Hybrid) | ~104.58 | 0.46 | 8.30 | 1.58* |
| PQC Speedup vs. RSA | 99.39% faster | 78.9% faster | 98.8% faster | — |

Table III shows the comparative latency results of the RSA-2048, ML-KEM-768 and proposed DW-HKEM hybrid model.The comparative latency results of the RSA-2048, ML-KEM-768 and the proposed DW-HKEM hybrid model are shown in Table III. To assess the performance improvement and computational efficiency, key generation, encapsulation, decapsulation and total execution latency are calculated for 100 iterations.

* DW-HKEM total handshake latency (encapsulation and decapsulation stages, excluding key generation , a per-session cost).

The results confirm that the time to be spent on the generation of the key dominates the session initialization at 103.96 ms, which is the typical time to spend per session in typical deployments where key pairs are generated once, and then reused for multiple handshakes. The overhead of DW-HKEM is only 0.96ms per-handshake, which includes 0.18ms of ML-KEM operations and about 0.78ms of overhead that comes from the integration of the oqs.dll in python. It is not noticed by end users in interactive banking applications (human-perceptible latency threshold: >100ms) and it is not even noticed in high throughput servers. Financial regulators in some countries have imposed re-keying policies for financial transactions, such as TLS 1.3 forward secrecy, per-transaction key isolation and re-keying at high frequency, where the 99.39x acceleration of ML-KEM-768 key generation over RSA-2048 (0.63 ms vs. 103.96 ms) has direct practical implications. These trends are confirmed in the live line chart on the CryptoX dashboard (as shown in Fig. 1) during the test session, which displays a hybrid mode operating at 11.66 ms/op, which includes realistic end-to-end latency including Flask request handling, AES-256 operations and JSON serialization overhead.

### *C. Communication Overhead Analysis*

***Table IV: Communication Overhead — Payload Size Comparison (Bytes)***

| Object | RSA-2048 | ML-KEM-768 | DW-HKEM | Overhead vs. RSA |
|---|---|---|---|---|
| Public Key | 256 B | 1,184 B | 1,440 B | +463% |
| Ciphertext | 256 B | 1,088 B | 1,392 B | +444% |

| Object | RSA-2048 | ML-KEM-768 | DW-HKEM | Overhead vs. RSA |
|---|---|---|---|---|
| Combined Handshake | 512 B | 2,272 B | 2,832 B | +453% |
| Ciphertext vs. RSA | Baseline | ~425% larger | ~5,800% larger | — |

Communication overhead analysis of RSA-2048, ML-KEM-768 and the proposed DW-HKEM is presented in table IV. The evaluation compares the size of public keys, the size of the ciphertexts and the handshake overhead of all the public key, hybrid and post-quantum mechanisms combined to analyze how much communication overhead is introduced by post-quantum and hybrid cryptographic mechanisms.

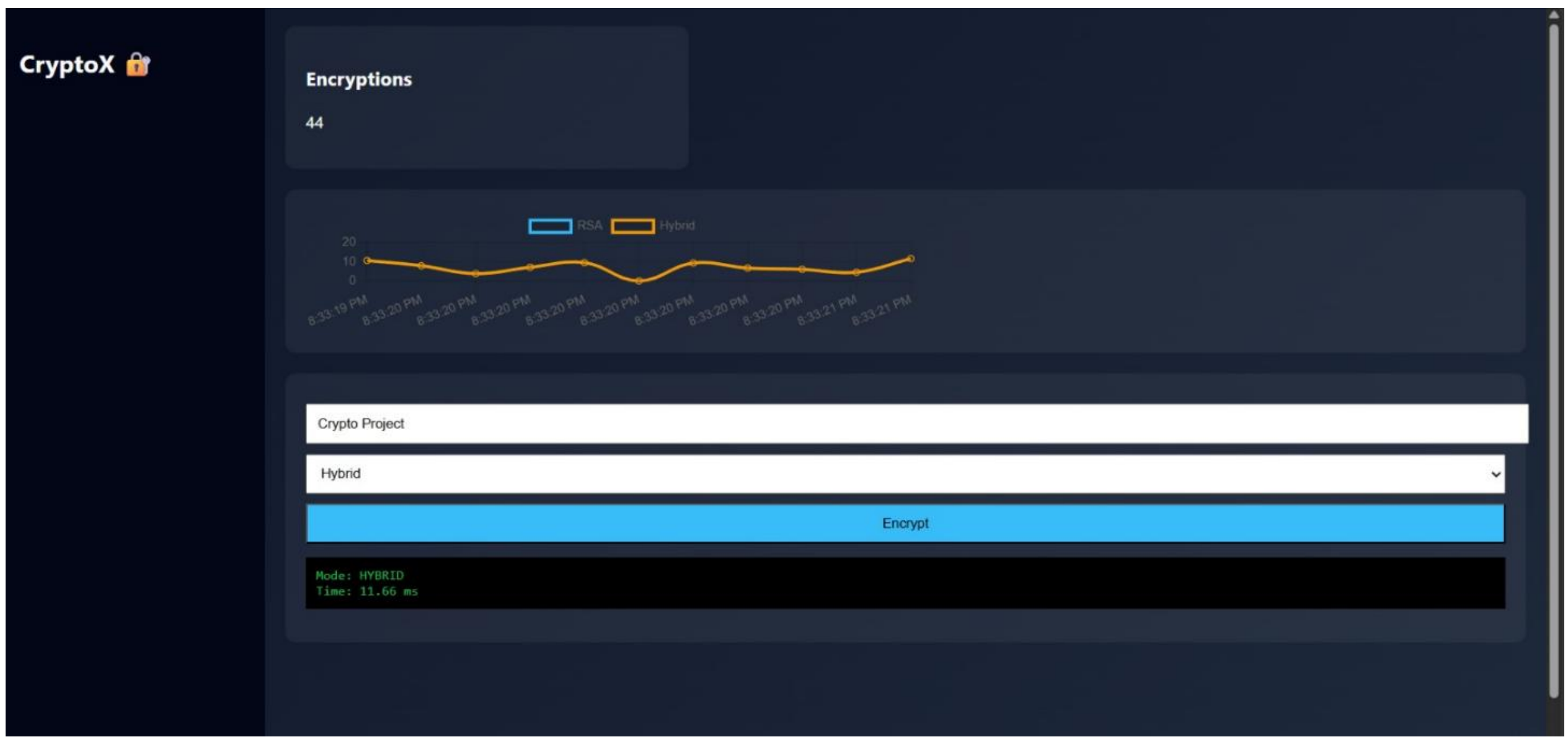

Fig. 1: CryptoX Real-Time Analytics Dashboard — live execution time trend chart (RSA and Hybrid modes) with operation count summary and encryption interface. Hybrid mode execution time: 11.66 ms measured in session.

Figure 1 illustrates the CryptoX real-time analytics dashboard displaying encryption execution trends for RSA and Hybrid cryptographic modes. The dashboard provides operational statistics, execution-time visualization, and live encryption session monitoring for performance analysis of the proposed hybrid cryptographic framework.

***Table V: Threat Analysis — Traditional RSA Banking vs. DW-HKEM Hybrid Protection***

| Security Aspect | Traditional RSA Banking System | Proposed DW-HKEM System |
|---|---|---|
| Quantum Resistance | None. RSA-2048 is broken by Shor's algorithm on a CRQC. Session keys recoverable. | Full. ML-KEM-768 layer protected by MLWE — conjectured quantum-intractable (FIPS 203). |

| Security Aspect | Traditional RSA Banking System | Proposed DW-HKEM System |
|---|---|---|
| MITM / Interception Protection | Encrypted channel only. Passive packet capture by adversary yields stored ciphertext for future decryption. | Dual-encapsulated session key. Even if RSA layer is intercepted, K_Session requires both S_RSA and S_KEM. |
| HNDL Resistance | None. Harvested RSA ciphertext fully decryptable once CRQC available. Historical transactions at risk. | Strong. Harvested ciphertext yields only S_RSA; K_Session remains secure while MLWE holds. |
| Real-Time Monitoring | Absent. Encryption operates as an opaque black-box. No performance baseline or anomaly detection. | Integrated. CryptoX dashboard provides live per-operation timing, trend charts, and session statistics. |
| Session Key Security | Single-primitive RSA. One algorithmic break compromises all session keys. | Dual-primitive KDF: K_Session = SHA-256(S_RSA ‖ S_KEM). Requires simultaneous break of RSA and MLWE. |
| Cryptographic Visibility | Zero. No instrumentation of encryption events, execution latency, or operational throughput. | Full. Every encrypt/decrypt event is timestamped, logged, and rendered live on the analytics dashboard. |
| Operational Analytics | Not available. Security architects cannot assess migration cost or validate SLA compliance. | Available. RSA vs. Hybrid trend lines enable direct overhead comparison and SLA validation pre-deployment. |
| Future Readiness | None. Full protocol replacement required; no quantum-safe fallback layer in current architecture. | Designed for migration. RSA layer maintains backward interoperability while ML-KEM provides quantum resilience. |
| Financial Compliance Alignment | FIPS 186 (RSA) only. Does not satisfy emerging NIST PQC migration guidance for financial infrastructure. | NIST FIPS 203 (ML-KEM) compliant. Positions deploying institutions ahead of anticipated PQC regulatory mandates. |

Table V compares the traditional RSA based banking system with the proposed DW-HKEM hybrid protection banking system to analyze the threats. The comparison assesses the resistance to quantum attacks, security of the session key, protection against attack by the interception of the process, capabilities of operational analytics, and future-proof capabilities, highlighting the security benefits and practical advantages of the suggested hybrid cryptographic system.

Figure 2 illustrates the banking threat scenario flow demonstrating a passive HNDL (Harvest Now, Decrypt Later) interception attack on a traditional RSA-protected banking session. The diagram shows how an attacker can silently capture encrypted TLS/RSA traffic during transmission and store the ciphertext for future decryption using cryptographically relevant quantum computers (CRQCs).

Figure 3 presents the life cycle of a Harvest Now, Decrypt Later (HNDL) attack on an RSA encrypted banking communication. The diagram shows the three stages of the attack and how the passive traffic capturing is followed by long-term encryption of the data stored, and how the data will be decrypted at a later time with a cryptographically relevant quantum computer (CRQC), which can perform Shor's algorithm.

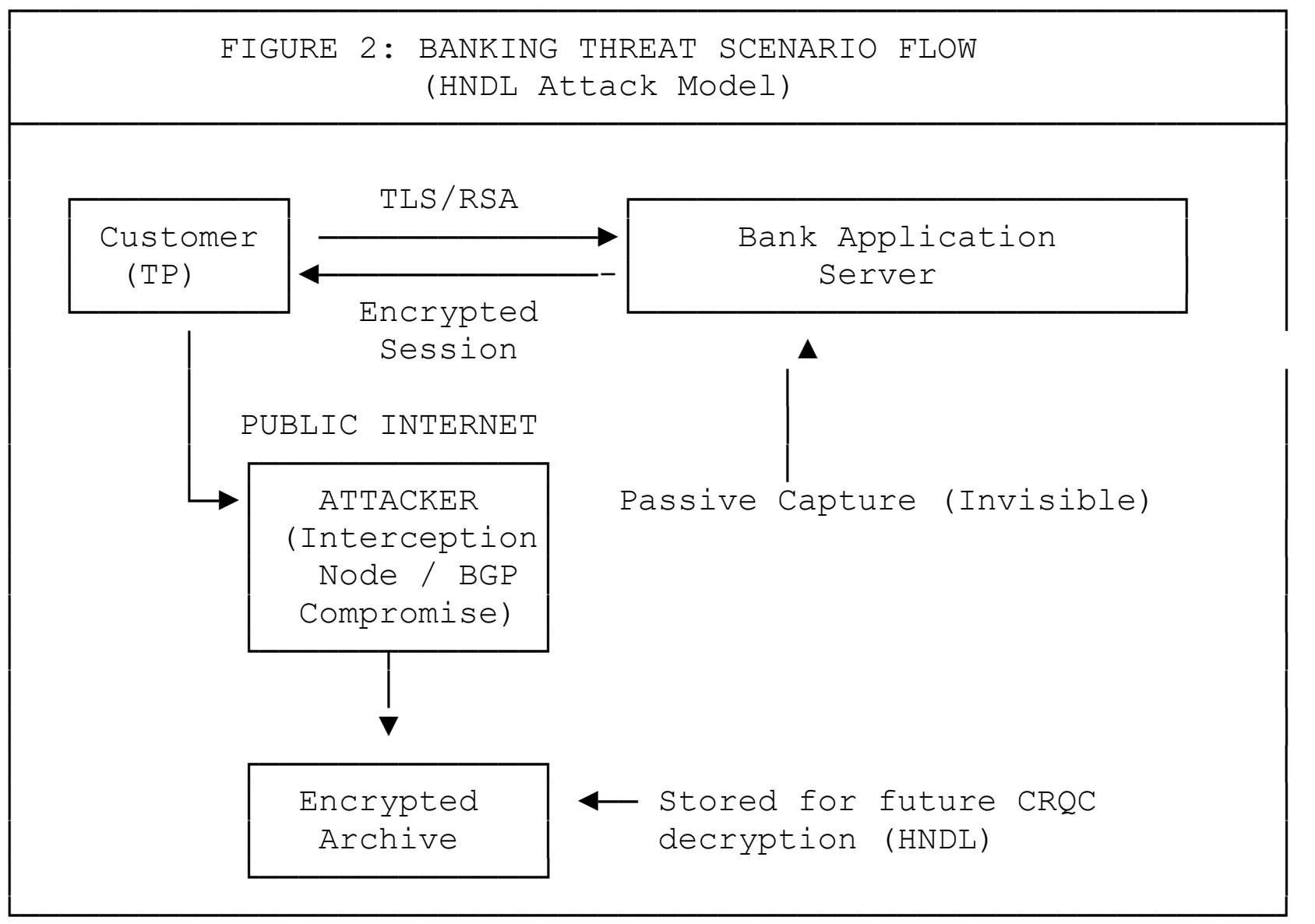

*Fig. 2. Banking Threat Scenario Flow — Passive HNDL Interception of RSA-Protected Banking Session.*

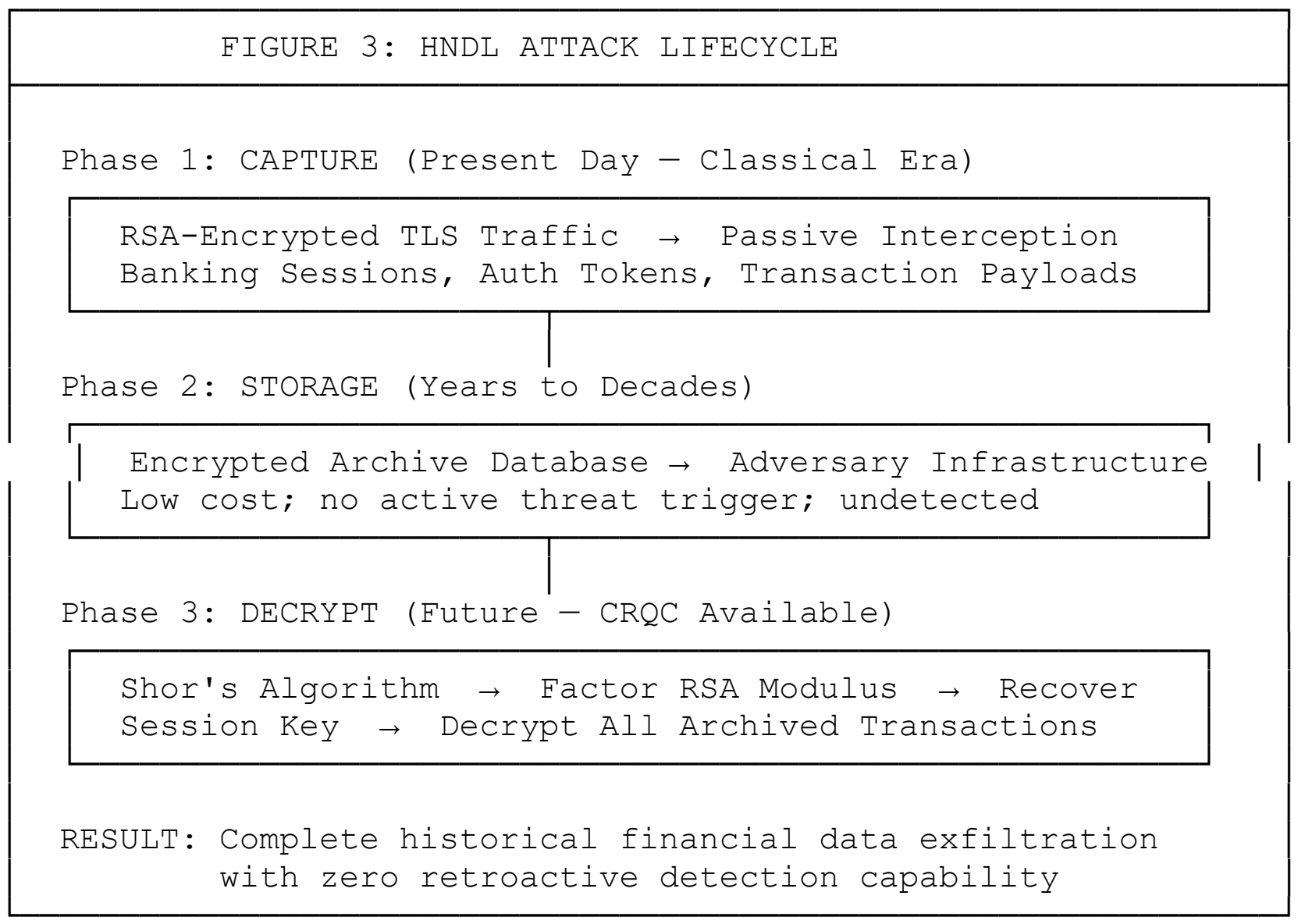

*Fig. 3. HNDL Attack Lifecycle — Three-Phase Harvest, Store, and Retrospective Decryption Timeline.*

An architecture for banking protection based on the combination of RSA-2048 with ML-KEM-768 is proposed in Figure 4, within a hybrid cryptographic framework. The architecture includes two layers of session key encapsulation, key derivation using the SHA-256 cryptographic function, and AES-256 transaction encryption that will provide quantum-resistant protection in secure banking communications and transaction processing.

The real-time monitoring dashboard flow for the CryptoX architecture designed for cryptographic observability and operational analytics is shown in Figure 5. The workflow illustrates the flow of encryption events to the metrics engine, through the REST API backend, visualization dashboard for live execution-time monitoring, session statistics, anomaly detection support and SLA validation in banking transaction environments.

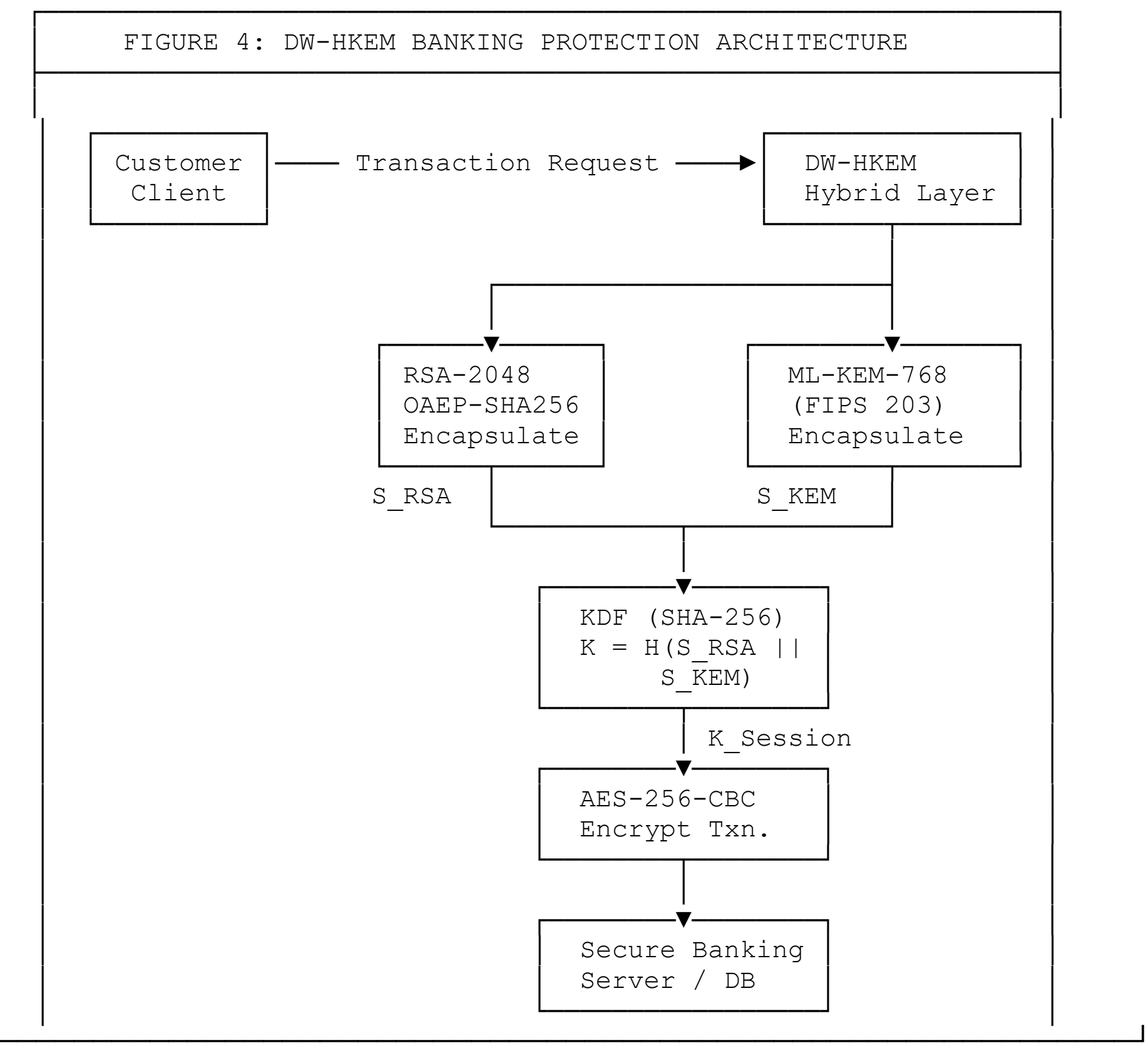

*Fig. 4. Proposed DW-HKEM Banking Protection Architecture — Dual-Layer Session Key Encapsulation with SHA-256 KDF Derivation.*

## VI. SECURITY AND PERFORMANCE ANALYSIS

### *A. Combinatorial Security Guarantee*

This is the Combinatorial Security Guarantee.This is the Combinatorial Security Guarantee. The construction of DW-HKEM is a more strictly secure than either of RSA-2048 or ML-KEM-768. Under three different threat models the hash K_Session = SHA-256(S_RSA || S_KEM) is computationally indistinguishable from a uniformly random 256-bit string. The classical adversary has a computational security of 2048-bit for RSA-2048 and 178-bit for ML-KEM-768 and the hybrid is at least as secure as the stronger primitive. In the presence of a quantum adversary running a CRQC that has Shor capabilities: RSA is broken, S_RSA is recovered, but K_Session is still secure as long as S_KEM is still computationally hidden (protected by MLWE). The security backstop is provided by the RSA layer in the hypothetical future classical adversary has a vulnerability in the MLWE layer.The security backstop is given by the RSA layer in the hypothetical future classical adversary had a vulnerability in the MLWE layer. The dual-wrap construction has this special security feature: the guarantee that the data can be transmitted both ways.

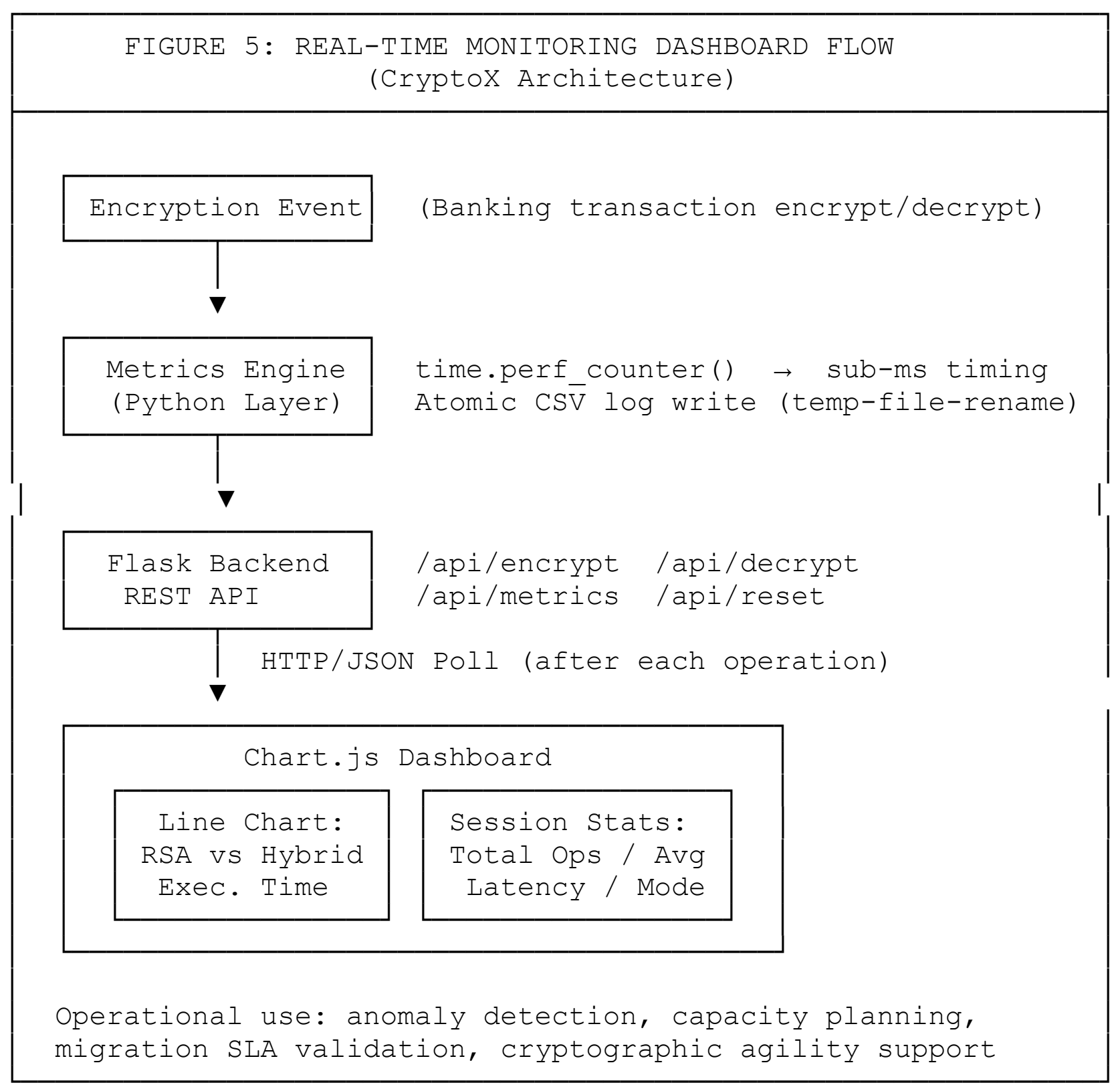

***Fig. 5. Real-Time Monitoring Dashboard Flow — CryptoX Event-to-Visualization Architecture for Cryptographic Observability.***

### *B. Banking-Sector Security Implications and Financial Compliance*

Acute threat model for financial institutions with the HNDL. Financial transaction data must be protected for longer periods of time as per Banking Regulations in various jurisdictions such as PCI DSS (Payment Card Industry Data Security Standard) and RBI Cyber Security Guidelines, India. However, during the regulatory retention period when the account credentials that were harvested in an HNDL attack and stored by an adversary are supposed to be retained with the account for five to 10 years, they may be decryptable by CRQC.

The DW-HKEM architecture explicitly tackles this regulatory aspect. Institutions implementing DW-HKEM can confidently claim that sessions started in this architecture are not susceptible to quantum decryption in the future by any adversary that is bound by the MLWE assumption. This puts DW-HKEM in the forefront of the financial institutions' compliance journey, starting to consider NIST PQC migration deadlines.

The CryptoX monitoring dashboard offers an added complimentary compliance: operational evidence of cryptographic systems performance and integrity. Regulators are increasingly looking for not just to use the right algorithms, but to show that they are being actively monitored when running crypto operations. A dashboard that generates a per-operation execution log with timestamps and trends, with the ability to see deviations, and statistics at the session level is a real-time audit-ready financial regulatory requirement that can be met by a cryptographic operational log.

### *C. Quantum-Safe Transaction Protection and Fraud Prevention*

DW-HKEM's guarantee of quantum-safe security applies in a not-trivial way when it comes to preventing fraud. Session authentication tokens can be intercepted, and if they are decryptable, can be replayed or used to set up fraudulent transactions after the end of the legitimate session in today's financial fraud scenarios. RSA-only protection means that an authentication token archive (collected by an HNDL) is a fraud instrument as soon as a CRQC is available. While the HNDL model has a deferred fraud vector due to the encryption of authentication tokens with the K_Session key, this is dealt with by the computational inaccessibility of the tokens when they are encrypted with a K_Session key under DW-HKEM protection.

In addition, CryptoX's real-time performance monitoring enables active fraud detection right at the cryptographic level. Large deviations from normal execution times (latency, number of operations, throughput) in a move that are inconsistent with other modes may be a sign of a cryptographic downgrade attack on the banking session establishment layer, tampering of the library in use, or a compromise of the HSM. Indicators are only visible in black-box deployments when a live performance dashboard is calibrated to the baseline expected in the environment, and are not visible otherwise.

### *D. Why the Hybrid System is Marginally Slower*

The one handshake overhead is 0.96ms against pure RSA due to two reasons at DW-HKEM. The first — ML-KEM encapsulation and decapsulation (0.08 + 0.10 = 0.18 ms) — is an intrinsic part of the second cryptographic primitive, and is the minimal cost of quantum-secure key establishment. The second (~0.78 ms) is due to calls to oqs.dll via the ctypes/cffi interface from Python-C. This bridge overhead is not algorithmic but implementation-specific: When deployed in production code in C or Rust, the cost of the bridge is negligible, leaving the overall hybrid overhead at around 0.18ms, which is not noticeable in any financial transaction environment.

### *E. The Migration Tax: Bandwidth vs. Computation*

The conclusion of the performance analysis is that the quantum migration cost is mainly due to the bandwidth penalty and not to the computational cost. The additional transmission latency caused by the 5,800% ciphertext expansion (1.392 B compared to 256 B) contributes about 0.009 milliseconds to a transmission latency of 1G bps enterprise network (negligible for any financial infrastructure application). In a TLS 1.3 handshake, there is a computational overhead of only 0.96ms, which is within the tolerances. But constraints of financial endpoints, such as Zigbee (IEEE 802.15.4, 127 byte MAC frame) or LoRaWAN (51-242 byte uplink), pose real challenges for DW-HKEM deployments, to which a protocol level fragmentation or ML-KEM-512 parameter downgrade would be necessary.

### *F. Real-Time Monitoring: Operational Relevance for Financial Deployments*

A monitoring layer in production financial services would have three different operational functions, similar to CryptoX. First, the performance anomaly detection: The execution time trend chart gives an operational baseline in cryptography which shows as soon as the performance deviates from it, due to resource exhaustion, algorithm downgrade attacks, or library compromise. Second, capacity planning: trends in the operation counts tell us important rotation frequencies, scaling the number of workers, and the size of the session cache – all critical parameters in a financial transaction processing environment where a known number of transactions per second is required. Third, migration validation: simultaneous migration validation of both RSA and hybrid timing information allows security architects to validate the hybrid overhead against SLA requirements, before full migration to the hybrid environment, reducing the risks of SLA compliance failures due to performance concerns during migration to the hybrid environment.

### *G. Scalability Considerations*

The in-memory operation log and single-threaded Flask development server in the CryptoX system themselves are limiting features of the system when it comes to scalability, fitting for the demonstration setting. When there is an operational load from multiple financial transactions, shared key objects and the operation log would need synchronization primitives. In a banking environment, production deployment would involve getting rid of the development server and replace it with a WSGI server (Gunicorn with multiple worker processes) and WebSocket-based push updates instead of fetch-after-operation polling, and persistent log storage would be replaced with either SQLite or PostgreSQL. Because the DW-HKEM gateway is stateless in all points other than session key pair, it can be horizontally scaled without change to the architecture and can support load-balanced deployments of the banking applications running on application servers.

## VII. CONCLUSION

In the proposed work, a Hybrid Key Encapsulation Mechanism (HKEM) named Dual-Wrap Hybrid Key Encapsulation Mechanism, (DW-HKEM), has been introduced for classical and quantum security, which combines two keys, namely, the 2048-bit RSA and 768-bit ML-KEM, using a SHA-256 based key derivation process. This future architecture was evaluated by implementing it and running 100 iterations of the benchmarks to measure the computational and communication overheads in a representative financial transaction environment.

The study showed that a hybrid post-quantum deployment can be implemented with very little computations overhead and still be protected from the threat model: Harvest-Now-Decrypt-Later (HNDL). A real time monitoring platform, CryptoX, was also built to provide operational visibility into cryptographic performance, with access to real time analytics, latency monitoring and session level statistics.

The key achievements of this work are: (i) Design and implementation of a combination of the two key post-quantum crypto-systems, namely RSA-2048 and ML-KEM-768, for a DW-HKEM framework; (ii) Empirical performance analysis of the hybrid architecture; (iii) Analysis of the risks of HNDL in banking environments and; (iv) Development of a real-time cryptographic observability dashboard for the post-quantum migration planning and operational monitoring.

Source Code: GitHub: https://github.com/sidhu1401sp-cpu/Cryptography-Project